\documentclass[onecolumn,aps,showpacs,showkeys,preprintnumbers,amsmath,amssymb,nofootinbib,superscriptaddress]{revtex4}
\usepackage{epsfig}
\usepackage{amsfonts}
\usepackage{dcolumn}

\newcommand{\newc}{\newcommand}
\newc{\gsim}{\lower.7ex\hbox{$\;\stackrel{\textstyle>}{\sim}\;$}}
\newc{\lsim}{\lower.7ex\hbox{$\;\stackrel{\textstyle<}{\sim}\;$}}
\newc{\gev}{\,{\rm GeV}}
\newc{\mev}{\,{\rm MeV}}
\newc{\ev}{\,{\rm eV}}
\newc{\kev}{\,{\rm keV}}
\newc{\tev}{\,{\rm TeV}}

\newc{\mz}{M_Z}
\newc{\mpl}{M_*}
\newc{\mw}{m_{\rm weak}}
\newc{\nr}[1]{N^c_R{}_{#1}}
\usepackage{amsmath}

\def\beq{\begin{equation}}
\def\eeq{\end{equation}}
\def\bea{\begin{eqnarray}}
\def\eea{\end{eqnarray}}
\def\bi{\begin{itemize}}
\def\ei{\end{itemize}}
\newc{\ie}{{\it i.e.}}          \newc{\etal}{{\it et al.}}
\newc{\eg}{{\it e.g.}}          \newc{\etc}{{\it etc.}}
\newc{\cf}{{\it c.f.}}

\def\inv{^{\raise.15ex\hbox{${\scriptscriptstyle -}$}\kern-.05em 1}}
\def\lbar{{\lower.35ex\hbox{$\mathchar'26$}\mkern-10mu\lambda}} 

\let\<=\langle
\let\>=\rangle

\let\+=\uparrow
\let\al=\alpha
\let\be=\beta
\let\ga=\gamma

\let\de=\delta

\let\ep=\epsilon

\let\la=\lambda

\let\th=\theta

\def\ph{\varphi}

\begin{document}
\author{Anders Basb\o ll}
\email{ab421@sussex.ac.uk} \affiliation{Department of Physics \& Astronomy, University of Sussex, Brighton, BN1 9QH, United Kingdom}
\title{$\nu$MSSM superpotential to 6th order - normalised and with no superfluous couplings}
\date{30th November, 2009}


\begin{abstract}
We expand the superpotential of $\nu$MSSM to 6th order. This is the order at which all flat directions can be lifted. All 5179 couplings are independent ie. the superpotential cannot be zero for all fields, without all couplings being zero. Likewise, any gauge invariant potential to the 6th order can be made by fixing the constants. A specific and welldefined choice of normalisation has been adopted. The case for investigating this potential, rather than looking at one or several generalised flat directions is made.
\end{abstract}
\maketitle \pagebreak

\section{Introduction}
The scalar potential of the Minimal Supersymmetric Standard Model
(MSSM) consists of F-terms ($\sum_{\phi}|\partial W/\partial \phi|^2$) and D-terms ($\sum_{a}g_a^2/2\left(\sum_{\phi}\phi^\dagger T^a \phi\right)^2$) where $T^a, W$ are the gauge generators and the superpotential respectively - and $\phi$ are the scalar fields. The potential has a large number of D-flat directions \cite{GheKolMar,Enqvist:2003gh}. Some of these are also F-flat when one considers the renormalisable superpotential only. If one allows higher order nonrenormalisable terms (only accepting R-parity as conserved) all these flat directions are lifted by different terms of different order. In this paper the superpotential is expanded to 6th order -- with reasonably normalised gauge invariant superfield products - and no superfluous couplings.

 The cosmological role of flat directions have been studied intensively\cite{Affleck:1984fy,Linde:1985gh,Dine:1995uk,Allahverdi:2006iq,Allahverdi:2006wh,Olive:2006uw,Allahverdi:2006xh,Allahverdi:2006gralep,us,me,Allahverdi:2008,GuiMetRioRiva,Olive08,warsaw,xflat,bjorn,RioRiva,AllDutMaz,brand,dufaux,KasKaw,ShoeKuse,Gumruk}. The issues include the possibility of the flat directions creating the baryon assymmetry of the universe, the flat direction vacuum
expectation values ($VEV$s) as a possible delayer of thermalisation or even the flat direction being the inflaton.
\section{A flat direction}

 The renormalisable part of the superpotential is \cite{Aitchison} $W_{renorm}=y_u^{ij}U_iQ_jH_u-y_d^{ij}D_iQ_j H_d-y_e^{ij}E_iL_jH_d+\mu H_uH_d$\footnote{Superfields and their scalar part will be represented by the same symbol. $Q,L,E,U,D,H_u,H_d$ are lefthanded quarks, lefthanded leptons, righthanded charged leptons, up-type righthanded quarks, down-type righthanded quarks, positive-hypercharged Higgs, negative-hypercharged Higgs.}
where the Yukawa couplings are the same as in the Standard Model\footnote{We will choose the basis where the SUSY-breaking mass terms - not the Yukawas - are diagonal.}.

Flatness means that the potential is zero for nonzero field values - which can be seen to happen if, and only if, all D-terms and F-terms vanish individually.  This only happens for exact (unbroken) SUSY and without nonrenormalisable terms in the superpotential\footnote{Adding a single soft term or a single  nonrenormalisable term will not lift all flat directions. But adding either all possible soft terms or all possible nonrenormalisable terms will. It would be very odd to allow for some soft terms and not others, or some nonrenormalisable terms, but not others. The spirit of the literature in general and \cite{Dine:1995uk,GheKolMar} in particular are that all terms allowed by the gauge symmetry and R-parity should present.}.

\subsection{Evolution of flat directions}
FD evolution was studied in \cite{Dine:1995uk}\footnote{QLD was the example.}\footnote{This section follows \cite{ABParis} closely.}. Giving $VEV$s to
\begin{equation}
 L_1^{1} =L_2^{2}=E_3=\phi/\sqrt{3}
\end{equation}
The superfields can be multiplied and the product can be parameterised by a canonical field for which we can write an equation of motion (including Hubble friction)
\begin{equation}
 \chi=L_1^{1}L_2^{2}E_3,\textbf{  } \chi=c\phi^m\textbf{  }(m=3),\textbf{  }\ddot{\phi}+3H\phi+V'(\phi)=0
\end{equation}
$m$ is used to keep example as general as possible. Adding nonrenormalisable terms to the superpotential ($M$ is a breaking scale: Planck/GUT/other) 
\begin{equation}
W=W_{renorm}+\sum_{n>3}\frac{\la}{M^{n-3}}\phi^n
\end{equation}
where all possible gauge invariant and R-parity conserving terms will be allowed (and expected to be of order 1). All FDs can be lifted by such terms - either by itself $\frac{\la}{nM^{n-3}}\chi=\frac{\la}{nM^{n-3}}\phi^n$ ($n=m$ - if positive R-parity), itself squared $\frac{\la}{nM^{n-3}}\chi^2=\frac{\la}{nM^{n-3}}\phi^n$ ($n=2m$ - if negative R-parity) or a combination of fields in the flat direction with exactly one field not in the direction $\frac{\la}{M^{n-3}}\psi\phi^{n-1}$ (with respect to which the derivative then can be taken). The potential is:
 \begin{equation}\label{FDE}
 V(\phi)=m_\phi^2|\phi|^2-cH^2|\ph|^2+\left(\frac{(Am_{3/2}+aH)\la \phi^n}{nM^{n-3}}+h.c\right)+|\la|^2\frac{|\phi|^{2n-2}}{M^{2n-6}}
\end{equation}
The masses are the ``real'' masses, whereas the terms of the same order in $\ph$ are Hubble induced terms. The F-term (last term) has order $2n-2$ because it is the derivative squared of n'th order ($W_n$ contributes to $V_{2n-2}$). The A-term (third term) is a coupling only between scalars. 

\subsection{Flatness vs. monomials}
Sometimes flatness is described by monomials. The relation is best illustrated by an example: $L_1L_2E_3=(\nu_e\mu-e\nu_\mu)\tau^c$ is a gauge invariant monomial. This gives flatness to either term ($\nu_e,\mu,\tau^c$) or ($e,\nu_\mu,\tau^c$) which can be $\phi$($e^{i\th_1},e^{i\th_2},e^{i\th_3}$) which keeps $D_a=0$ for all generators. Superterm $W_4\propto L_1L_2E_3N_1$ gives $|F_{N_1}|^2\propto|L_1L_2E_3|^2$ which is positive and thus lifts flatness, whereas the A-term is $A*e^{i*\th_A}L_1L_2E_3+h.c.$ which is negative for one of the mentioned field combinations and thus chooses the minimum -- if ``All fields equal to zero'' is not the minimum. This can easily happen during inflation, where the mass terms are negligible, if c is positive \cite{Dine:1995uk} or the A-term large compared to c \cite{KasKaw}. The minimum will be \cite{Dine:1995uk} $\left(\frac{\be HM^{n-3}}{\la}\right)^{\frac{1}{n-2}}$ where $\be$ is a numerical constant depending on $a$,$c$,$n$ and presumeably of order unity. Since $H$ can be very large during inflation, we can get very close to the breaking scale - especially for large values of $n$. A more detailed discussion of $LLE$ and flatness including gauge choices and parameterisation was given in \cite{mitkatalog}.

\subsection{The case for investigating the potential}
There are problems with this picture\footnote{these points made in \cite{ABParis}}. Monomials (or directions) are not independent. There are only 17 mass  terms -- or 20 if righthanded neutrinos($N$s) are included -- yet there are 712 (715 including $N$s)  independent monomials \cite{mitkatalog}. To illustrate, $m_{(QQQ)_\textbf{4}L_1L_2L_3E_1}^2=1/7(m_{Q_1}^2+m_{Q_2}^2+m_{Q_3}^2+m_{L_1}^2+m_{L_2}^2+m_{L_3}^2+m_{E_1}^2)$ while $m_{(QQQ)_\textbf{4}L_1L_2L_2E_1}^2=1/7(m_{Q_1}^2+m_{Q_2}^2+m_{Q_3}^2+m_{L_1}^2+2*m_{L_2}^2+m_{E_1}^2)$. These are clearly not independent. Also, if$(QQQ)_\textbf{4}L_1L_2L_3E_1$ has a VEV, so has $(QQQ)_\textbf{4}L_1L_2L_2E_1$. Also, when is $QQQLLLE$ broken? To fit the formula one could imagine that without $N$s it would be broken by itself squared - dimension 14. However, the space of $Q,L,E$ is 27(18+6+3) dimensional. It breaks the Standard Model completely, so D-terms remove 12 complex degrees of freedom (c.d.o.f.)\footnote{One real non-flat direction and one real gauge choice for each.}. So the D-flat space is 15 dimensional. $W_4$ (4th order superpotential) includes $QQQL$ and $QULE$ -- so $F_Q,F_L,F_U,F_E$ give 36 complex constraints and thus $W_4$ lifts the flat direction. This means it is lifted by the 6th order in the potential- eventhough its A-term is of much higher order. Including $N$s will give A-terms like $QQQLLLEN$ but the direction will still be lifted by $W_4$ (including $LLEN$).

Also, the point that the F-term has no phase dependence is not correct once a superfield is involved in more than one monomial i.e. even in the renormalisable part $F_{H_d}$ has $QD$ and $LE$ terms \cite{GheKolMar}. Another example of this is given in \cite{AllDutMaz}.

The point is, that the potential for a flat direction is actually only given by \ref{FDE} if all other couplings than the A-term and W-term for the monomial itself is put to zero. This is a very strong condition to impose - and if doing so, one should have to argue for it. In other words, the full potential must be considered.

\subsection{A word of caution}
In this paper the superpotential of $\nu MSSM$ is expanded to 6th order - the order at which all flatness would be lifted \cite{mitkatalog}. However, the 6th order of the superpotential corresponds to the 10th order in the potential. This means that for a consistent expansion in the potential, if one includes 6th order in $W$, one should really include direct scalar couplings to order 10. That is a massive task\footnote{A very rough counting suggest something of the order 2 million couplings.}.  

\section{Notation}
Color indices are $a,b...$, anticolor $\overline{a},\overline{b}...$. Family indices are latin letters $i,j,k...$, $SU(2)_L$ are greek letters $\al ,\be, \ga...$, ($\de , \ep$ are reserved for the Kronecker delta and maximally antisymmetric tensors) and other latin letters ($\textbf{e}$,$\textbf{f}$) in bold are used occassionally to mark other dimensionalities -- that is to mark the different between different products made of identical superfields. Summation over repeated indices is implied if and only if one is lowered and the other is raised. 

Once monomials are combined, the symmetries are quite different depending on to which extent family indices of the same superfield is contracted. For instance, $H_uL+LLE$ is quite different depending on whether there is a repeated $L$-field (only one way to contract) or no repeated $L$-field (3 ways to contract - but the 3 sum to zero). Therefore the products will be categorised by their participating superfields - not the monomials from which they can be made. So we name $H_uL2LE$ and $H_uLLLE$. This is also convenient since they need different indices. The former needs 2 indices for $L$s and one for $E$, whereas the latter needs only one for $E$ and one for which contraction is made. To make indices more efficient, for $L2L$ we use one index (i) for the generation of $L2$ and a second ($\textbf{j}$) restricted to values ${1,2}$ such that the generation of $L$ is $i+\textbf{j}$. $\textbf{j}$ is in bold to show that is not a generational index, but it does have values in the same range in order to make the summation of indices clear. Generational indices are modulo 3 [more precisely: gen(j)=Modulo(j-1,3)+1]. For every combination of superfields, the indices start from left to right, and for superfields that appear more than once the indices are given as: All generations appearing to the same power: no index. Two generations appearing to the same power (different from the third): One index - the generation of the third, regardless of whether its power is higher or lower than the others. All generations with different powers: 2 indices. One for the generation of the highest power, and one for the index of the middle power compared to the first index - as in the $L2L$ example above.

\section{Normalisation}
Normalisation of the couplings is clearly an issue when one is considering writing arbitrary couplings. Here are the choices made in this paper: $\ep$ antisymmetric tensors used in gauge contractions are normalised by one. A repeated superfield is normalised by $1/n!$ where $n$ is the number of time it occurs. To make the reading easy, in the tables below these facors are written for each superfield. For instance, $Q2L2D2$ is normalised by $1/(2!2!2!)$. The normalisation has to be done by reappering superfield, not reappearing field, since, for instance, $(H_uH_d)^2=(H^+H^-)^2+2H^+H^-H_u^0H_d^0+(H_u^0H_d^0)^2$ so the reappearence of fields differs for the same coupling. When some linear combination of products are zero, we have chosen a linear algebra approach such that if, say, $a,b,c$ are products (including gauge contractions) of superfields and $a+b+c=0$ then $(a-b)/\sqrt{2}$ and $(a+b-2c)/\sqrt{6}$ are used as basis vectors. Using all three of them will overcount the number of couplings. Does this single out $c$? No, not really. Thinking of $a=(-1/2,\sqrt{3}/2)$, $b=(-1/2$, $-\sqrt{3}/2)$, $c=(1,0)$ which sum to zero, one finds $(0,\sqrt{3/2})$ and $(-\sqrt{3/2},0)$ as the new basis vectors. This looks as if $c$ is kept, though boosted. However, regardless which vectors you assign indices $a,b,c$ you will always get two orthogonal vectors of length $\sqrt{3/2}$ which makes the normalisation consistent -- we have an orthogonal basis. Furthermore, letting $a,b,c$ be normally distributed aroud zero for each coordinate, will give a variance of (3/2,3/2) exactly the same as if given normal distributions to coordinates in the new basis. This means, at least for normal distribution functions, the probabilities will all be the same regardless of whether one uses the 3 original vectors or the 2 new one - but the latter highlights that there is, in fact, only 2 degrees of freedom. If one should have a theory concerning one or more of the original vectors, it is easy to translate into the new ones. 

\section{Neutrinos}
We start with righthanded neutrinos, $N$s. In table \ref{Ntabel} all the neutrino products are listed. The dimension and number od complex degrees of freedom (c.d.o.f.) is given. The neutrinos, having no gauge quantum numbers, only have symmetries among themselves. That is, adding neutrinos to other directions is quite simple. It is just an outer product of the neutrino and non-neutrino parts. If the non neutrino direction itself has negative R-parity, you add every odd number of neutrinos. If it has positive R-parity, neutrinos are added in even quantities. 

\begin{table}[b]
\caption{\label{Ntabel} Neutrino products.}
\begin{ruledtabular}
\begin{tabular}{lllllll}
Name &Expression & Dimension & c.d.o.f. & R-parity\\
\hline
$(N)_i $ & $N_i, [1\leq i\leq 3]$& $1$ & $3$ & $-$ \\
$(N2)_i $ & $N_i^2/2!, [1\leq i\leq 3]$& $2$ & $3$ & $+$ \\
$(NN)_i $ & $N_{i+1}*N_{i+2}, [1\leq i\leq 3]$& $2$ & $3$ & $+$ \\
$(N3)_i $ & $N_i^3/3!, [1\leq i\leq 3]$& $3$ & $3$ & $-$ \\
$(N2N)_{i,\textbf{j}} $ & $N_{i}^2*N_{i+\textbf{j}}/2!, [1\leq i\leq 3,1\leq \textbf{j}\leq 2]$& $3$ & $6$ & $-$ \\
$(NNN) $ & $N_{1}*N_{2}*N_3$& $3$ & $1$ & $-$ \\
$(N4)_i $ & $N_i^4/4!, [1\leq i\leq 3]$& $4$ & $3$ & $+$ \\
$(N3N)_{i,\textbf{j}} $ & $N_{i}^3*N_{i+\textbf{j}}/3!, [1\leq i\leq 3,1\leq \textbf{j}\leq 2]$& $4$ & $6$ & $+$ \\
$(N2N2)_i $ & $N_{i+1}^2*N_{i+2}^2/(2!2!), [1\leq i\leq 3]$& $4$ & $3$ & $+$ \\
$(N2NN)_{i} $ & $N_{i}^2*N_{i+1}*N_{i+2}/2!, [1\leq i\leq 3]$& $4$ & $3$ & $+$ \\
$(N5)_i $ & $N_i^5/5!, [1\leq i\leq 3]$& $5$ & $3$ & $-$ \\
$(N4N)_{i,\textbf{j}} $ & $N_{i}^4*N_{i+\textbf{j}}/4!, [1\leq i\leq 3,1\leq \textbf{j}\leq 2]$& $5$ & $6$ & $-$ \\
$(N3N2)_{i,\textbf{j}} $ & $N_{i}^3*N_{i+\textbf{j}}^2/(3!2!), [1\leq i\leq 3,1\leq \textbf{j}\leq 2]$& $5$ & $6$ & $-$ \\
$(N3NN)_{i} $ & $N_{i}^3*N_{i+1}*N_{i+2}/3!, [1\leq i\leq 3]$& $5$ & $3$ & $-$ \\
$(N2N2N)_i $ & $N_i*N_{i+1}^2*N_{i+2}^2/(2!2!), [1\leq i\leq 3]$& $5$ & $3$ & $-$ \\
$(N6)_i $ & $N_i^6/6!, [1\leq i\leq 3]$& $6$ & $3$ & $+$ \\
$(N5N1)_{i,\textbf{j}} $ & $N_{i}^5*N_{i+\textbf{j}}/(5!), [1\leq i\leq 3,1\leq \textbf{j}\leq 2]$& $6$ & $6$ & $+$ \\
$(N4N2)_{i,\textbf{j}} $ & $N_{i}^4*N_{i+\textbf{j}}^2/(4!2!), [1\leq i\leq 3,1\leq \textbf{j}\leq 2]$& $6$ & $6$ & $+$ \\
$(N4NN)_{i} $ & $N_{i}^4*N_{i+1}*N_{i+2}/4!, [1\leq i\leq 3]$& $6$ & $3$ & $-$ \\
$(N3N3)_i $ & $N_{i+1}^3*N_{i+2}^3/(3!3!), [1\leq i\leq 3]$& $6$ & $3$ & $+$ \\
$(N3N2N)_{i,\textbf{j}} $ & $N_{i}^3*N_{i+\textbf{j}}^2*N_{i-\textbf{j}}/(3!2!), [1\leq i\leq 3,1\leq \textbf{j}\leq 2]$& $6$ & $6$ & $+$ \\
$(N2N2N2) $ & $N_{1}^2*N_{2}^2*N_3^2/(2!2!2!)$& $6$ & $1$ & $+$ \\
\end{tabular}
\end{ruledtabular}
\end{table}
\section{The tables of couplings}
Here follows the tables of couplings. For all products one takes a coupling constant and divide (or multiply) by the breaking scale to the correct order ie. taking one example from each dimension in table \ref{Monomialtabel}: $W=\la_{H_uH_d}(H_uH_d)*M+...+\la_{QH_dD}^{i,j}(QH_dD)_{i,j}+....+\la_{LLEN}^{i,j,k}(LLE)_{i,j}*N_k/M+....+\la_{DDDLH_d}^{i}(DDDLH_d)_{i}/M^2+......+\la_{QLDN2N}^{i,j,k,l,\textbf{m}}(QLD)_{i,j,k}*(N2N)_{l,\textbf{m}}/M^3$ or, using the full expressions $W=\la_{H_uH_d}H_u^{\al}H_d^{\be}\ep_{\al \be}*M+...\la_{QH_dD}^{i,j}Q_i^{\al ,a}H_d^{\be}D_j^{\overline{a}}\ep_{\al \be}\de_{a\overline{a}}+....+\la_{LLEN}^{i,j,k}L_{i+1}^{\al}L_{i+2}^{\be}E_j\ep_{\al \be}N_k/M+....+\la_{DDDLH_d}^{i}D_1^{\overline{a}}D_{2}^{\overline{b}}D_{3}^{\overline{c}}L_{i}^{\al}H_d^{\be}\ep_{\overline{a}\overline{b}\overline{c}}\ep_{\al \be}/M^2+......+\la_{QLDN2N}^{i,j,k,l,\textbf{m}}Q_i^{\al ,a}L_j^{\be}D_k^{\overline{a}}\ep_{\al \be}\de_{a\overline{a}}N_{l}^2*N_{l+\textbf{m}}/2!/M^3$. Full expressions are given in the tables, except of the always trivial manouvre of adding neutrinos to other combinations. In the tables the c.d.o.f. of the expression is given, aswell as the c.d.o.f. after neutrinos have been added. Neutrinos are only added in numbers to give positive R-parity. Thus the 3rd line of table \ref{Monomialtabel} should be read as: $H_L$ is dimension 2 and has 3 c.d.o.f. It has negative R-parity in itself - marked by the number being in bracket. Adding one righthanded neutrino gives dimension 3 and has 9 c.d.o.f. Adding three gives dimension 5 and 30 c.d.o.f.
\subsection{Monomials}
In table \ref{Monomialtabel} the monomials are presented. $Q$s are the the most involved superfields due to their participation in both $SU(3)$ and $SU(2)$ contractions. So here are some definitions used. All the monomials where spelled out in detail in \cite{mitkatalog} from which these definitions also are taken. $(QQQ)_\textbf{4}^{\al \be \ga}=Q_1^{\al a}Q_2^{\be b}Q_3^{\ga c}\ep_{abc}\ep^{ijk}/\sqrt{6}$ where the $\textbf{4}$ denotes that the $QQQ$s transform as a $\textbf{4}$ under $SU(2)_L$. The $\textbf{2}$s are first defined by $(QQQ)_{ijk}^{\al}=Q_{i}^{a \be}Q_{j}^{b \ga}Q_{k}^{a \al}\ep_{abc}\ep_{\be \ga}$, then $Q2Q_{i\textbf{j}\textbf{h}}=\frac{QQQ_{i,i+j,i}+QQQ_{i+j,i,i}-2QQQ_{i,i,i+j}}{\sqrt{6}}$ and $QQQ_{\textbf{i}}= (QQQ_{i+1,i+2,i}+QQQ_{i+2,i+1,i})/\sqrt{2}$ and then   $QQQ_{\textbf{7}}=(QQQ_{\textbf{1}}-QQQ_{\textbf{2}})/\sqrt{2}$ and $QQQ_{\textbf{8}}=(QQQ_{\textbf{1}}+QQQ_{\textbf{2}}-2QQQ_{\textbf{3}})/\sqrt{6}$. 

Also $(Q2Q2)_{i\textbf{d}}=(Q2Q)_{i+1,1\textbf{h}}Q_{i+2}-(Q2Q)_{i+2,2\textbf{h}}Q_{i+1}/\sqrt{2}$,  
and then
$Q2QQ_{i\textbf{7}}^a=(Q2Q_{i,1\textbf{h}}^\al Q_{i+2}^{\be,a} \ep_{\al\be}-Q2Q_{i,2\textbf{h}}^\al Q_{i+1}^{\be,a} \ep_{\al\be})/\sqrt{2}$, $Q2QQ_{i\textbf{8}}^a=(QQQ_\textbf{i+1}-QQQ_\textbf{i+2})^\al Q_i^{\be,a} \ep_{\al\be}/\sqrt{2}$ and $Q2QQ_{i\textbf{9}}^a=(2QQQ_\textbf{i}-QQQ_\textbf{i+1}-QQQ_\textbf{i+2})^\al Q_i^{\be,a} \ep_{\al\be}/\sqrt{12}+(Q2Q_{i,1\textbf{h}}^\al Q_{i+2}^{\be,a} \ep_{\al\be}+Q2Q_{i,2\textbf{h}}^\al Q_{i+1}^{\be,a} \ep_{\al\be})/2$.

\begin{table}[b]
\caption{\label{Monomialtabel}Monomials. -- means wrong $R$-parity or lower dimension than the entry, () means c.d.o.f. of entry, but wrong $R$-parity. For each dimension, the c.d.o.f. is listed. In the first line, the indices must sum to the number of the dimension.}
\begin{ruledtabular}
\begin{tabular}{lllllll}
Name &Expression & Dim 2 & Dim 3& Dim 4& Dim 5& Dim 6\\
\hline
$(N)_i$ summary of table \ref{Ntabel} & $N_1^{n1}N_2^{n2}N_3^{n3}/(n1!n2!n3!)$   & $\textbf{6}$& (10) &$\textbf{15}$ &(21) &$\textbf{28} $\\
$(H_uH_d) $ & $H_u^{\al}H_d^{\be}\ep_{\al \be}$   & $\textbf{1}$& -- &$\textbf{6}(2N)$ &-- &$\textbf{15}(4N) $\\
$(H_uL)_i$ &$ H_u^{\al}L_i^{\be}\ep_{\al \be},[1\leq i\leq 3]$   &$ (3)$&$\textbf{9}(N)$&-- &$\textbf{30}(3N)$ &-- \\
$(QH_uU)_{i,j}$ &$ Q_i^{\al ,a}H_u^{\be}U_j^{\overline{a}}\ep_{\al \be}\de_{a\overline{a}},[1\leq i,j\leq 3]$  &-- &$\textbf{9}$&--&$\textbf{54}(2N)$&--  \\
$(QH_dD)_{i,j}$ &$ Q_i^{\al ,a}H_d^{\be}D_j^{\overline{a}}\ep_{\al \be}\de_{a\overline{a}},[1\leq i,j\leq 3]$   &--&$\textbf{9}$&--&$\textbf{54}(2N)$&--  \\
$(LH_dE)_{i,j}$ &$ L_i^{\al}H_d^{\be}E_j\ep_{\al \be},[1\leq i,j\leq 3]$  &-- &$\textbf{9}$&--&$\textbf{54}(2N)$&--  \\
$(QLD)_{i,j,k}$ &$ Q_i^{\al ,a}L_j^{\be}D_k^{\overline{a}}\ep_{\al \be}\de_{a\overline{a}},[1\leq i,j,k\leq 3]$  &-- &(27)&$\textbf{81}(N)$&--&$\textbf{270}(3N)$ \\
$(LLE)_{i,j}$ &$ L_{i+1}^{\al}L_{i+2}^{\be}E_j\ep_{\al \be},[1\leq i,j\leq 3]$   &--&(9)&$\textbf{27}(N)$&--&$\textbf{90}(3N)$ \\
$(UDD)_{i,j}$ &$ U_i^{\overline{a}}D_{j+1}^{\overline{b}}D_{j+2}^{\overline{c}}\ep_{\overline{a}\overline{b}\overline{c}},[1\leq i,j\leq 3]$ &-- &(9)&$\textbf{27}(N)$&--&$\textbf{90}(3N)$ \\
$(UUDE)_{i,j,k} $ & $U_{i+1}^{\overline{a}}U_{i+2}^{\overline{b}}D_{j}^{\overline{c}}E_k\ep_{\overline{a}\overline{b}\overline{c}},[1\leq i,j,k\leq 3]$&-- &--& $\textbf{27}$& -- & $\textbf{162}(2N)$\\
$(QLUE)_{i,j,k,l} $ & $Q_i^{\al ,a}L_j^{\be}U_k^{\overline{a}}E_l\ep_{\al \be}\de_{a\overline{a}},[1\leq i,j,k,l\leq 3]$&-- &--& $\textbf{81}$& -- & $\textbf{486}(2N)$\\
$(Q2UD)_{i,j,k} $ & $Q_{i}^{\al a}U_{j}^{\overline{a}}Q_{i}^{\be b}D_{k}^{\overline{b}}\de_{a\overline{a}}\de_{b\overline{b}}\ep_{\al \be}/2!,[1\leq i,j,k\leq 3]$&-- &--& $\textbf{27}$& -- & $\textbf{162}(2N)$\\
$(QQUD)_{i,j,k,\textbf{e}} $ & $Q_{i+\textbf{e}}^{\al a}U_{k}^{\overline{a}}Q_{i-\textbf{e}}^{\be b}D_{l}^{\overline{b}}\de_{a\overline{a}}\de_{b\overline{b}}\ep_{\al \be},[1\leq i,j,k\leq 3,1\leq \textbf{e}\leq 2]$&-- &--& $\textbf{54}$& -- & $\textbf{324}(2N)$\\
$(QH_dUE)_{i,j,k}$&$Q_i^{\al ,a}H_d^{\be}U_j^{\overline{a}}E_k\ep_{\al \be}\de_{a\overline{a}},[1\leq i,j,k\leq 3]$&-- &--&(27)&$\textbf{81}(N)$&--\\
$(Q2QL)_{i,\textbf{j},k}$&$QQQ_{i\textbf{j}\textbf{h}}^{\al}L_{k}^{\be}\ep_{\al \be}/2!,[1\leq i,k\leq 3,1\leq\textbf{j}\leq 2]$
&-- &--& $\textbf{18}$& -- & $\textbf{108}(2N)$\\
$(QQQL)_{\textbf{e},k}$&$QQQ_{\textbf{e}}^{\al}L_{k}^{\be}\ep_{\al \be},[1\leq k\leq 3,7\leq e\leq 8]$&-- &--& $\textbf{6}$& -- & $\textbf{36}(2N)$\\
$(Q2QH_d)_{i,\textbf{j}}$&$QQQ_{i\textbf{j}\textbf{h}}^{\al}H_d^{\be}\ep_{\al \be}/2!,[1\leq i\leq 3,1\leq\textbf{j}\leq 2]$
&-- &--& (6)& $\textbf{18}(N)$&--\\
$(QQQH_d)_{\textbf{e}}$&$QQQ_{\textbf{e}}^{\al}H_d^{\be}\ep_{\al \be},[7\leq e\leq 8]$&-- &--& 
 (2)& $\textbf{6}(N)$&--\\
$(DDDLH_d)_{i}$&$D_1^{\overline{a}}D_{2}^{\overline{b}}D_{3}^{\overline{c}}L_{i}^{\al}H_d^{\be}\ep_{\overline{a}\overline{b}\overline{c}}\ep_{\al \be},[1\leq i\leq 3]$& -- & --& -- &$\textbf{3}$ & --\\
$(DDDLL)_{i}$&$D_1^{\overline{a}}D_{2}^{\overline{b}}D_{3}^{\overline{c}}L_{i+1}^{\al}L_{i+2}^{\be}\ep_{\overline{a}\overline{b}\overline{c}}\ep_{\al \be},[1\leq i\leq 3]$&--&--&--&(3)& $\textbf{9}(N)$\\
$(QQUUE)_{i,j\textbf{e},k}$&$Q_{i+1}^{\al a}U_{j+\textbf{e}}^{\overline{a}}Q_{i+2}^{\be b}U_{j-\textbf{e}}^{\overline{b}}E_k\de_{a\overline{a}}
\de_{b\overline{b}}\ep_{\al \be},$&--&--&--&(54)& $\textbf{162}(N)$\\
&$[1\leq i,j,k\leq 3,1\leq \textbf{e}\leq 2]$&&&&\\
$(Q2UUE)_{i,j,k}$&$Q_{i}^{\al a}U_{j+1}^{\overline{a}}Q_{i}^{\be b}U_{j+2}^{\overline{b}}E_k\de_{a\overline{a}}
\de_{b\overline{b}}\ep_{\al \be}/2!,$&--&--&--&(27)& $\textbf{81}(N)$\\
&$[1\leq i,j,k\leq 3]$&&&&\\
$(QQU2E)_{i,j,k}$&$Q_{i+1}^{\al a}U_{j}^{\overline{a}}Q_{i+2}^{\be b}U_{j}^{\overline{b}}E_m\de_{a\overline{a}}
\de_{b\overline{b}}\ep_{\al \be}/2!,$&--&--&--&(27)& $\textbf{81}(N)$\\
&$[1\leq i,j,k\leq 3]$&&&&\\
$(UUUE2)_{i}$&$U_1^{\overline{a}}U_{2}^{\overline{b}}U_{3}^{\overline{c}}E_iE_j\ep_{\overline{a}\overline{b}\overline{c}}/2!,[1\leq i\leq 3]$&--&--&--&(3)& $\textbf{9}(N)$\\
$(UUUEE)_{i}$&$U_1^{\overline{a}}U_{2}^{\overline{b}}U_{3}^{\overline{c}}E_{i+1}E_{i+2}
\ep_{\overline{a}\overline{b}\overline{c}},[1\leq i\leq 3]$&--&--&--&(3)& $\textbf{9}(N)$\\
$(QQQ_{\textbf{4}}LH_uH_d)_{i}$&$(QQQ)_{\textbf{4}}^{\al \be \ga}L_i^{\al '} H_u^{\be '} H_d^{\ga '}\ep _{\al \al '}\ep _{\be \be '}\ep _{\ga \ga '},[1\leq i\leq 3]$&--&--&--&--& $\textbf{3}$\\
$Q2Q2U_{i,j}$&$(Q2Q2)_{i\textbf{d}}^a U_j^{\overline{a}} \de_{a\overline{a}}/(2!2!)[1\leq i,j\leq 3]$&--&--&--&(9)& $\textbf{27}(N)$\\
$Q3QU_{i,\textbf{j},k}$&$(Q2Q_{i\textbf{j}\textbf{h}}^{\al} Q_i^{\be a} U_k^{\overline{a}} \ep_{\al \be}\de_{a\overline{a}}/3!,$&--&--&--&(18)& $\textbf{54}(N)$\\
&$[1\leq i,k\leq 3,1\leq \textbf{j}\leq 2]$&&&&\\
$Q2QQU_{i,\textbf{e},j}$&$QQQQ_{i\textbf{e}}^a U_j^{\overline{a}}\de_{a\overline{a}}/2!,$&--&--&--&(27)& $\textbf{81}(N)$\\
&$[1\leq i,j\leq 3,7\leq e\leq 9]$&&&&\\
\end{tabular}
\end{ruledtabular}
\end{table}

\subsection{Trivial products}
In table \ref{TriPro} are the trivial products i.e. there is no other way to contract indices of the participating superfields than to make the contractions of the participating monomials. The only thing to do is to multiply by symmetry factor for repeated superfields.

\begin{table}[b]
\caption{\label{TriPro}Trivial products. -- means wrong $R$-parity or lower dimension than the entry, () means c.d.o.f. of entry, but wrong $R$-parity. For each dimension, the c.d.o.f. is listed.}
\begin{ruledtabular}
\begin{tabular}{lllllll}
Name &Expression & Dim 2 & Dim 3& Dim 4& Dim 5& Dim 6\\
\hline
$H_u2H_d2 $ & $(H_uH_d)*(H_uH_d)/2!$   & --& -- &$\textbf{1}$ &-- &$\textbf{6}(2N)$\\
$(H_u2H_dL)_i $ & $(H_uH_d)*(H_uL)_{i}/2!,[1\leq i\leq 3]$   & --& -- &$(3)$&$\textbf{9}(N)$ &-- \\
$(H_uH_d2LE)_{i,j} $ & $(H_uH_d)*(LH_dE)_{i,j}/2!,[1\leq i,j\leq 3]$   & --& --& -- &$\textbf{9}$ &-- \\
$(H_uH_d2QD)_{i,j} $ & $(H_uH_d)*(QH_dD)_{i,j}/2!,[1\leq i,j\leq 3]$   & --& --& -- &$\textbf{9}$ &-- \\
$(H_uH_d2QU)_{i,j} $ & $(H_uH_d)*(QH_uU)_{i,j}/2!,[1\leq i,j\leq 3]$   & --& --& -- &$\textbf{9}$ &-- \\
$(UDDH_uL)_{i,j,k} $ & $(UDD)_{i,j}*(H_uL)_{k},[1\leq i,j,k\leq 3]$   & --& --& -- &$\textbf{27}$ &-- \\
$(UDDH_uH_d)_{i,j} $ & $(UDD)_{i,j}*(H_uH_d),[1\leq i,j\leq 3]$   & --& --& -- &(9) &$\textbf{27}(N)$ \\
$(QUH_u2L)_{i,j,k} $ & $(QH_uU)_{i,j}*(H_uL)_{k}/2!,[1\leq i,j,k\leq 3]$   & --& --& -- &(27) &$\textbf{81}(N)$ \\
$(QDH_d2LE)_{i,j,k,l} $ & $(QH_dD)_{i,j}*(LH_dE)_{k,l}/2!,[1\leq i,j,k,l\leq 3]$   & --& --& -- &-- &$\textbf{81}$ \\
$(UDDLLE)_{i,j,k,l} $ & $(UDD)_{i,j}*(LLE)_{k,l},[1\leq i,j,k,l\leq 3]$   & --& --& -- &-- &$\textbf{81}$ \\
$(UUDEH_uH_d)_{i,j,k} $ & $(UUDE)_{i,j,k}*(H_uH_d),[1\leq i,j,k\leq 3]$   & --& --& -- &-- &$\textbf{27}$ \\
$H_u3H_d3 $ & $(H_uH_d)^3/6!$   & --& -- &-- &-- &$\textbf{1}$\\
\end{tabular}
\end{ruledtabular}
\end{table}
\subsection{Symmetry only between monomials}
In table \ref{MonSym} are the directions that are trivial products i.e. there is no other way to contract indices of the participating superfields than make the contractions of the participating monomials. The only thing to do is to multiply by symmetry factor for repeated superfields.

Here the notation of repeated superfields is perhaps a disadvantage -- after all $QLD+QLD$ is just an outer product (except for permutation factors). It has been checked for this, and all the other products, that there are no reductions to be made ie. $Q_1L_1D_1*Q_2L_2D_2$, $Q_1L_1D_2*Q_2L_2D_1$  and the 6 other products with the same superfields cannot make a linear combination to zero.

\begin{table}[b]
\caption{\label{MonSym}Symmetry only between monomials. -- means wrong $R$-parity or lower dimension than the entry, () means c.d.o.f. of entry, but wrong $R$-parity. For each dimension, the c.d.o.f. is listed.}
\begin{ruledtabular}
\begin{tabular}{lllllll}
Name &Expression & Dim 2 & Dim 3& Dim 4& Dim 5& Dim 6\\
\hline
$(H_u2L2)_{i} $ & $(H_uL)_{i}^2/(2!2!),[1\leq i \leq 3]$   & --& -- &$\textbf{3}$ &-- &$\textbf{18}(2N)$\\
$(H_u2LL)_{i} $ & $(H_uL)_{i+1}*(H_uL)_{i+2}/(2!),[1\leq i \leq 3]$   & --& -- &$\textbf{3}$ &-- &$\textbf{18}(2N)$\\
$(H_u3L2H_d)_{i} $ &$ (H_uL)_{i}^2(H_uH_d)/(3!2!),[1\leq i\leq 3]$   & --& -- &-- &-- &$\textbf{3}$\\
$(H_u3LLH_d)_{i} $ &$ (H_uL)_{i+1}*(H_uL)_{i+2}*(H_uH_d)/(3!),[1\leq i\leq 3]$   & --& -- &-- &-- &$\textbf{3}$\\
$(Q2H_d2D2)_{i,j} $ & $(QH_dD)_{i,j}^2*/(2!2!2!),[1\leq i,j\leq 3]$   & --& -- &-- &-- &$\textbf{9}$\\
$(Q2H_d2DD)_{i,j} $ & $(QH_dD)_{i,j+1}*(QH_dD)_{i,j+2}/(2!2!),[1\leq i,j\leq 3]$   & --& -- &-- &-- &$\textbf{9}$\\
$(QQH_d2D2)_{i,j} $ & $(QH_dD)_{i+1,j}*(QH_dD)_{i+2,j}/(2!2!),[1\leq i,j\leq 3]$   & --& -- &-- &-- &$\textbf{9}$\\
$(QQH_d2DD)_{i,j,\textbf{e}} $ & $(QH_dD)_{i+1,j+\textbf{e}}*(QH_dD)_{i+2,j-\textbf{e}}/(2!),[1\leq i,j\leq 3, 1\leq \textbf{e}\leq 2]$   & --& -- &-- &-- &$\textbf{18}$\\
$(Q2H_u2U2)_{i,j} $ & $(QH_uU)_{i,j}^2*/(2!2!2!),[1\leq i,j\leq 3]$   & --& -- &-- &-- &$\textbf{9}$\\
$(Q2H_u2UU)_{i,j} $ & $(QH_uU)_{i,j+1}*(QH_dD)_{i,j+2}/(2!2!),[1\leq i,j\leq 3]$   & --& -- &-- &-- &$\textbf{9}$\\
$(QQH_u2U2)_{i,j} $ & $(QH_uU)_{i+1,j}*(QH_uU)_{i+2,j}/(2!2!),[1\leq i,j\leq 3]$   & --& -- &-- &-- &$\textbf{9}$\\
$(QQH_u2UU)_{i,j,\textbf{e}} $ & $(QH_uU)_{i+1,j+\textbf{e}}*(QH_uU)_{i+2,j-\textbf{e}}/(2!),[1\leq i,j\leq 3,1\leq \textbf{e}\leq 2]$   & --& -- &-- &-- &$\textbf{18}$\\
$(Q2L2D2)_{i,j,k} $ & $(QLD)_{i,j,k}^2/(2!2!2!),$   & --& -- &-- &-- &$\textbf{27}$\\
&$[1\leq i,j,k \leq 3]$&&&&\\
$(Q2L2DD)_{i,j,k} $ & $(QLD)_{i,j,k+1}*(QLD)_{i,j,k+2}/(2!2!),$   & --& -- &-- &-- &$\textbf{27}$\\
&$[1\leq i,j,k\leq 3]$&&&&\\
$(Q2LLD2)_{i,j,k} $ & $(QLD)_{i,j+1,k}*(QLD)_{i,j+2,k}/(2!2!),$   & --& -- &-- &-- &$\textbf{27}$\\
&$[1\leq i,j,k\leq 3]$&&&&\\
$(QQL2D2)_{i,j,k} $ & $(QLD)_{i+1,j,k}*(QLD)_{i+2,j,k}/(2!2!),$   & --& -- &-- &-- &$\textbf{27}$\\
&$[1\leq i,j,k\leq 3]$&&&&\\
$(Q2LLDD)_{i,j,k,\textbf{e}} $ & $(QLD)_{i,j+1,k+\textbf{e}}*(QLD)_{i,j+2,k-\textbf{e}}/(2!),$   & --& -- &-- &-- &$\textbf{54}$\\
&$[1\leq i,j,k\leq 3, 1\leq\textbf{e}\leq 2]$&&&&\\
$(QQL2DD)_{i,j,k,\textbf{e}} $ & $(QLD)_{i+1,j,k+\textbf{e}}*(QLD)_{i+2,j,k-\textbf{e}}/(2!),$   & --& -- &-- &-- &$\textbf{54}$\\
&$[1\leq i,j,k\leq 3, 1\leq\textbf{e}\leq 2]$&&&&\\
$(QQLLD2)_{i,j,k,\textbf{e}} $ & $(QLD)_{i+1,j+\textbf{e},k}*(QLD)_{i+2,j-\textbf{e},k}/(2!),$   & --& -- &-- &-- &$\textbf{54}$\\
&$[1\leq i,j,k\leq 3, 1\leq\textbf{e}\leq 2]$&&&&\\
$(QQLLDD)_{i,j,k,\textbf{e},\textbf{f}} $ & $(QLD)_{i+1,j+\textbf{e},k+\textbf{f}}*(QLD)_{i+2,j-\textbf{e},k-\textbf{f}},$   & --& -- &-- &-- &$\textbf{108}$\\
\end{tabular}
\end{ruledtabular}
\end{table}
\subsection{2 $E$s}
In table \ref{EEtabel} are the products with 2 $E$s. The $E$'s are not part of any contractions, so their symmetries are always independent of the rest of the directions.  As with N's, their dimensionality can be calculated seperately. After all, they are the same as $N$s except for hypercharge. For instance 9 dimensional LHdE and 9 dimentional LLE is easily split into 6 dimentional EE (not 9) and LL and LHd up to 9 dimensions can be investigated seperately  (is in fact 8 dimensional - due to a 4 $SU(2)$ fields  doublet redundancy).

\begin{table}[b]
\caption{\label{EEtabel}2 $E$s -- means wrong $R$-parity or lower dimension than the entry, () means c.d.o.f. of entry, but wrong $R$-parity. For each dimension, the c.d.o.f. is listed.}
\begin{ruledtabular}
\begin{tabular}{lllllll}
Name &Expression & Dim 2 & Dim 3& Dim 4& Dim 5& Dim 6\\
\hline
$(L2H_d2E2)_{i,j} $ & $ (LH_dE)_{i,j}^2/(2!2!2!),$   & --& -- &-- &-- &$\textbf{9}$\\
&$[1\leq i\leq j,\leq 3]$&&&&\\
$(L2H_d2EE)_{i,j} $ & $ (LH_dE)_{i,j+1}*(LH_dE)_{i,j+2}/(2!2!),$   & --& -- &-- &-- &$\textbf{9}$\\
&$[1\leq i\leq j,\leq 3]$&&&&\\
$(LLH_d2E2)_{i,j} $ & $ (LH_dE)_{i+1,j}*(LH_dE)_{i+2,j}/(2!2!),$   & --& -- &-- &-- &$\textbf{9}$\\
&$[1\leq i\leq j,\leq 3]$&&&&\\
$(LLH_d2EE)_{i,j} $ & $ (LH_dE)_{i+1,j+1}*(LH_dE)_{i+2,j+2}/2!,$   & --& -- &-- &-- &$\textbf{9}$\\
&$[1\leq i\leq j,\leq 3]$&&&&\\
$(L2L2E2)_{i,j} $ & $ (LLE)_{i,j}^2/(2!2!2!),$   & --& -- &-- &-- &$\textbf{9}$\\
&$[1\leq i\,j\leq 3]$&&&&\\
$(L2L2EE)_{i,j} $ & $ (LLE)_{i,j+1}*(LLE)_{i,j+2}/(2!2!),$   & --& -- &-- &-- &$\textbf{9}$\\
&$[1\leq i\,j\leq 3]$&&&&\\
$(L2LLE2)_{i,j} $ & $(LLE)_{i+1,j}*(LLE)_{i+2,j}/(2!2!),$   & --& -- &-- &-- &$\textbf{9}$\\
&$[1\leq i\,j\leq 3]$&&&&\\
$(L2LLEE)_{i,j} $ & $(LLE)_{i+1,j+1}*(LLE)_{i+2,j+2}/(2!),$   & --& -- &-- &-- &$\textbf{9}$\\
&$[1\leq i,j\leq 3]$&&&&\\
\end{tabular}
\end{ruledtabular}
\end{table}
\subsection{4 $SU(2)$ Superfields}
When more than one combination of monomials contain the same superfields more care is needed. Products with 4 $SU(2)$ charged fields are in table \ref{SU2tabel}.  Take HuHd+LLE and HuL+LHdE (consisting of, say, $HuL_1+L_2HdE$ and $HuL_2+L_1HdE$). E can be seperated, so we have 4 superfields to contract with SU(2). If the 2 $L$s are of the same generation it is trivial - each of them contracting with a Higgs. If the $L$s are different we have 4 different superfields to contract. Changing to general notation, (A*B)(D*C)+(A*C)(B*D)+(A*D)(C*B)=0 \cite{GheKolMar}. One could use just 2 of the products to span the nonzero space, but it is more clear to write unitvectors that are perpendicular to the zerospace. Therefore we define the double valued function $SU(2)_\textbf{4}[A,B,C,D]= {[(A*B)(D*C)-(A*C)(B*D)]/\sqrt{2}, [(A*B)(D*C)+(A*C)(B*D)-2(A*D)(C*B)]/\sqrt{6}}$. Thus if the $L$s are the same there are 3*3 dimensions, if they are different there are 3*2*3 dimensions - altogether 27 dimensions. 

In fact, the same situation can arise, without there been seperate monomials. We define ``Pseudo''monomials   to mean what would have been monomials if hypercharge wasn't there. Take $LHu+QHdD$ and $HuHd+QLD$. While this looks straight forward as generations giving 27 combinations and then the 2 ways to contract - in fact, one can also make gauge invariant product $LH_d+QH_uD$ which is a combination of pseudomonomials. Thus, as before, there are 3 contractions, of which the  sum is zero. So there is still 27+27=54 dimentions, but the right way to get to it is not just to add the 2 products from basis monomials.

In many cases, the treatment is dependent on whether generations are repeated or not. Take $HuL+QDL$. If the $L$s are different, we have 4 doublets that can contract in any way (the 2 $L$s can contract) - and as usual this gives 2 degrees of freedom. If the $L$s are of the same generation, we just have a trivial product. The first case also gives the same number of degrees of freedom as the outer product - but it is built in another way.

\begin{table}[b]
\caption{\label{SU2tabel}4 $SU(2)$ Superfields . -- means wrong $R$-parity or lower dimension than the entry, () means c.d.o.f. of entry, but wrong $R$-parity. For each dimension, the c.d.o.f. is listed.}
\begin{ruledtabular}
\begin{tabular}{lllllll}
Name &Expression & Dim 2 & Dim 3& Dim 4& Dim 5& Dim 6\\
\hline
$(Q2UDH_uH_d)_{i,j,k,\textbf{e}} $ & $ SU(2)_\textbf{4}[Q_{i}^{a}U_{j}^{\overline{a}}\de_{a\overline{a}},
Q_{i}^{b}D_{k}^{\overline{b}}\de_{b\overline{b}}H_uH_d]_{\textbf{e}}/2!,$   & --& -- &-- &-- &$\textbf{54}$\\
&$[1\leq i,j,k\leq 3,1\leq \textbf{e}\leq 2]$&&&&\\
$(QQUDH_uH_d)_{i,j,k,\textbf{e},\textbf{f}} $ & $ SU(2)_\textbf{4}[Q_{i+\textbf{e}}^{a}U_{j}^{\overline{a}}\de_{a\overline{a}},
Q_{i-\textbf{e}}^{b}D_{k}^{\overline{b}}\de_{b\overline{b}}H_uH_d]_{\textbf{f}},$   & --& -- &-- &-- &$\textbf{108}$\\
&$[1\leq i,j,k\leq 3,1\leq \textbf{e},\textbf{f}\leq 2]$&&&&\\
$(QULEH_uH_d)_{i,j,k,\textbf{e},l} $ & $ SU(2)_\textbf{4}[Q_{i}^{a}U_{j}^{\overline{a}}\de_{a\overline{a}},
L_{k},H_u,H_d]_{\textbf{e}}E_l,$   & --& -- &-- &-- &$\textbf{162}$\\
&$[1\leq i,j,k,l\leq 3,1\leq \textbf{e}\leq 2]$&&&&\\
$(QDLH_uH_d)_{i,j,k,\textbf{e}} $ & $ SU(2)_\textbf{4}[Q_{i}^{a}D_{j}^{\overline{a}}\de_{a\overline{a}},
L_{k},H_u,H_d]_{\textbf{e}},$   & --& -- &-- &(54) &$\textbf{162}(N)$\\
&$[1\leq i,j,k\leq 3,1\leq \textbf{e}\leq 2]$&&&&\\
$(LLLH_uE)_{\textbf{e},i} $ & $ SU(2)_\textbf{4}[L_{1},L_{2},L_{3},H_u]_{\textbf{e}}E_{i},$   & --& -- &-- &$\textbf{6}$ &--\\
&$[1\leq i \leq 3,1\leq \textbf{e}\leq 2]$&&&&\\
$(L2LH_uE)_{i,\textbf{j},k} $ & $ (LLE)_{i+j,k}*(H_uL)_{i}/2!$   & --& -- &-- &$\textbf{18}$ &--\\
&$[1\leq i,k\leq 3,1\leq \textbf{j}\leq 2]$&&&&\\
$(QDLLH_u)_{i,j,k,\textbf{e}} $ & $ SU(2)_\textbf{4}[Q_{i}^{a}D_{j}^{\overline{a}}\de_{a\overline{a}},L_{k+1},L_{k+2},H_u]_{\textbf{e}},$   & --& -- &-- &$\textbf{54}$ &--\\
&$[1\leq i,j,k \leq 3,1\leq \textbf{e}\leq 2]$&&&&\\
$(QDL2H_u)_{i,j,k}  $ & $ (QLD)_{i,k,j}*(H_uL)_{k}/2!$   & --& -- &-- &$\textbf{27}$ &--\\
&$[1\leq i,j,k\leq 3]$&&&&\\
$(QDLLLE)_{i,j,k,\textbf{e}} $ & $ SU(2)_\textbf{4}[Q_{i}^{a}D_{j}^{\overline{a}}\de_{a\overline{a}},L_{1},L_{2},L_{3}]_{\textbf{e}}E_{k},$   & --& -- &-- &-- &$\textbf{54}$\\
&$[1\leq i,j,k \leq 3,1\leq \textbf{e}\leq 2]$&&&&\\
$(QDL2LE)_{i,j,k,\textbf{l},m}  $ & $ (QLD)_{i,k,j}*(LLE)_{k+\textbf{l}}*E_{m}/2!$   & --& -- &-- &-- &$\textbf{162}$\\
&$[1\leq i,j,k,m\leq 3,1\leq \textbf{l}\leq 2]$&&&&\\
$(H_uLLH_dE)_{i,\textbf{e},j} $ & $ SU(2)_\textbf{4}[H_u,L_{i+1},L_{i+2},H_d]_{\textbf{e}}E_{j},$   & --& -- &-- &(18) &$\textbf{54}(N)$\\
&$[1\leq i,j\leq 3,1\leq \textbf{e}\leq 2]$&&&&\\
$(H_uL2H_dE)_{i,j}  $ & $ (H_uL)_{i}*(LH_dE)_{i,j}/2!$   & --& -- &-- &(9) &$\textbf{27}(N)$\\
&$[1\leq i,j\leq 3]$&&&&\\
$((Q2Q)_2LH_uH_d)_{i,\textbf{j},k,\textbf{e}}$&$SU(2)_\textbf{4}[Q2Q_{i\textbf{j}\textbf{h}},L_{k},H_u,H_d]_{\textbf{e}}/2!,$&-- &--&--&--&$\textbf{36}$\\
&$[1\leq i,k\leq 3,1\leq \textbf{j},\textbf{e}\leq 2]$&&&&\\
$((QQQ)_2LH_uH_d)_{\textbf{e},i,\textbf{f}}$&$SU(2)_\textbf{4}[QQQ_{\textbf{e}},L_{i},H_u,H_d]_{\textbf{f}},$&-- &--&--&--&$\textbf{12}$\\
&$[1\leq i\leq 3,1\leq \textbf{e},\textbf{f}\leq 2]$&&&&
\end{tabular}
\end{ruledtabular}
\end{table}
At the last two entries of table \ref{SU2tabel} it might seem arbitrary that the $Q$s have been contracted first. However, $(QQQ)_4LH_uH_d$ was in table \ref{Monomialtabel}. Indeed, if one instead makes a general definition of independent directions from 6 SU(2) doublets ($SU(2)_\textbf{6}$), one finds, when using it on 3 $Q$s and 1 of each of $L,H_uH_d$ one finds no possibilities for all $Q$s of the same generation, whereas for one repeated Q index, one finds exactly the entries of the penultimate line in table \ref{SU2tabel}. For all $Q$s different, one finds the entries of the last line of table \ref{SU2tabel} aswell as $(QQQ)_4LH_uH_d$ - and the entries of the latter are orthogonal to the entries in the former.
\subsection{More than three $SU(3)$ fields}
Similar problems arise when there are more color contractions to be made. These products are in table \ref{SU3tabel}. $UDD+UDD$ seems simple (in itself it is in fact a trivial product) but it mixes with pseudomonomials $UUD+DDD$. The latter obviously demands the $U$s to be different and all 3 generations of $D$s to be present. In that case the sum of 3 terms (both combinations of U with DD and the DDD combination) vanished. This gives 2 free parameters, just as one would get if one ignored the $DDD$ possibility. But in order to keep the ``orthogonal to zero vectors'' approach, again we will take the combinations orthogonal to the zero combination. 

Also, $QLD+UDD$ can be made as $QLU+DDD$. Again, in itself $QLD+UDD$ is a trivial product. But the sum of the 3 combinations with all generations of D involved is equal to (minus) $QLU+DDD$ as they are defined in the text. 
\begin{table}[b]
\caption{\label{SU3tabel}More than 3 $SU(3)$ Superfields . -- means wrong $R$-parity or lower dimension than the entry, () means c.d.o.f. of entry, but wrong $R$-parity. For each dimension, the c.d.o.f. is listed.}
\begin{ruledtabular}
\begin{tabular}{lllllll}
Name &Expression & Dim 2 & Dim 3& Dim 4& Dim 5& Dim 6\\
\hline
$(U2D2D2)_{i,j} $ & $(UDD)_{i,j}^2/(2!2!2!),[1\leq i,j \leq 3]$   & --& -- &-- &-- &$\textbf{9}$\\
$(U2D2DD)_{i,j} $ & $(UDD)_{i,j+1}(UDD)_{i,j+2}/(2!2!),[1\leq i,j \leq 3]$   & --& -- &-- &-- &$\textbf{9}$\\
$(UUD2D2)_{i,j} $ & $(UDD)_{i+1,j}(UDD)_{i+2,j}/(2!2!),[1\leq i,j \leq 3]$   & --& -- &-- &-- &$\textbf{9}$\\
$(UUD2DD)_{i,j,\textbf{e}}, $& $[1\leq i,j,k\leq 3,7\leq \textbf{e} \leq 8]\textrm{ with }$& --& -- &-- &-- &$\textbf{18}$\\
$(UUD2DD)_{i,j,\textbf{e=7}} $ & $((UDD)_{i+1,j+1}(UDD)_{i+2,j+2})-(UDD)_{i+1,j+2}(UDD)_{i+2,j+1})/(2!\sqrt{2}),$ &&&&&\\
$(UUD2DD)_{i,j,\textbf{e=8}} $ &$((UDD)_{i+1,j+1}*(UDD)_{i+2,j+2}+(UDD)_{i+1,j+2}*(UDD)_{i+2,j+1}$   &&&&&\\
&$-2(UUD)_{i,j}*(DDD))/(2!\sqrt{6})$,&&&&&\\
$(QLUD2D)_{i,j,k,l,\textbf{m}} $ & $(UDD)_{k,l+\textbf{m}}*(QLD)_{i,j,l}/2!,$   & --& -- &-- &-- &$\textbf{162}$\\
&$[1\leq i,j,k,l \leq 3,1\leq \textbf{m} \leq 2]$ &&&&\\
$(QLUDDD)_{i,j,k,\textbf{e}} $ & $ with [1\leq i,j,k \leq 3,7\leq \textbf{e} \leq 9]$   & --& -- &-- &-- &$\textbf{81}$\\
$(QLUDDD)_{i,j,k,\textbf{e=7}} $ & $((UDD)_{k,1}*(QLD)_{i,j,1}-(UDD)_{k,2}*(QLD)_{i,j,2})/\sqrt{2},$   &&&&\\
& &&&&\\
$(QLUDDD)_{i,j,k,\textbf{e=8}} $ &  $((UDD)_{k,1}*(QLD)_{i,j,1}+(UDD)_{k,2}*(QLD)_{i,j,2}$  &&&&\\
&$-2(UDD)_{k,3}*(QLD)_{i,j,3})/\sqrt{6},$ &&&&\\
$(QLUDDD)_{i,j,k,\textbf{e=9}} $ &$ ((UDD)_{k,1}*(QLD)_{i,j,1}+(UDD)_{k,2}*(QLD)_{i,j,2}+,$ &&&&\\
&$(UDD)_{k,3}*(QLD)_{i,j,3}+3*(QLU)_{i,j,k}*(DDD)/\sqrt{12}$ &&&&
\end{tabular}
\end{ruledtabular}
\end{table}
with the obvious definitions $(UUD)_{i,j}=U_{i+1}^{\overline{a}}U_{i+2}^{\overline{b}}D_{j}^{\overline{c}}\ep_{\overline{a}\overline{b}\overline{c}}$,  $(DDD)=D_{1}^{\overline{a}}D_{2}^{\overline{b}}D_{3}^{\overline{c}}\ep_{\overline{a}\overline{b}\overline{c}}$ and $(QLU)_{i,j,k}=Q_i^{\al ,a}L_j^{\be}U_k^{\overline{a}}\ep_{\al \be}\de_{a\overline{a}},[1\leq i,j,k\leq 3]$.

\section{Conclusion and outlook}\label{scon}
We have expanded the $\nu MSSM$ superpotential to the 6th order. There are 5179 independent couplings. 7 of dimension 2, 36 of dimension 3, 376 of dimension 4, 468 of dimension 5 and 4292 of dimension 6. We have argued why it is neccessary to take this full potential into account rather than looking at generic flat directions when investigating the cosmological importence of flatness.
\section{Acknowledgements}
I would like to thank the Danish taxpayers who have supported this work through The Danish Council for Independent Research | Natural Sciences and I thank Stephan Huber, Mark Hindharsh and David Bailin for useful discussions.


\begin{thebibliography}{99}


\vspace{-2mm}

\bibitem{GheKolMar}
  T.~Gherghetta, C.~F.~Kolda and S.~P.~Martin,
  Nucl.\ Phys.\ B {\bf 468}, 37 (1996).

\bibitem{Enqvist:2003gh}
  K.~Enqvist and A.~Mazumdar,
  Phys.\ Rept.\  {\bf 380}, 99 (2003).
 
\bibitem{Affleck:1984fy}
  I.~Affleck and M.~Dine,
  Nucl.\ Phys.\  B {\bf 249}, 361 (1985).
  
\bibitem{Linde:1985gh}
  A.~D.~Linde,
  Phys.\ Lett.\  B {\bf 160}, 243 (1985).
 
\bibitem{Dine:1995uk}
  M.~Dine, L.~Randall and S.~D.~Thomas,
  Phys.\ Rev.\ Lett.\  {\bf 75}, 398 (1995).
  

\bibitem{Allahverdi:2006gralep}
  R.~Allahverdi and A.~Mazumdar,
   J.\ Cosmol.\ Astropart.\ Phys.  {\bf 0610}, 008 (2006).


\bibitem{Allahverdi:2006wh}
  R.~Allahverdi and A.~Mazumdar,
  Phys.\ Rev.\ D {\bf 76}, 103526 (2007).
 


\bibitem{Allahverdi:2006iq}
 R.~Allahverdi, K.~Enqvist, J.~Garcia-Bellido and A.~Mazumdar,
  Phys.\ Rev.\ Lett.\  {\bf 97} 191304 (2006).
 
 
\bibitem{KasKaw}
S.~Kasuya, M.~Kawasaki,
Phys.\ Rev.\ D {\bf 74}, 063507 (2006).


\bibitem{Olive:2006uw}
  K.~A.~Olive and M.~Peloso,
  Phys.\ Rev.\ D {\bf 74}, 103514 (2006).
  
\bibitem{Allahverdi:2006xh}
  R.~Allahverdi and A.~Mazumdar,
  J.\ Cosmol.\ Astropart.\ Phys. 08 (2007) 023.
  
\bibitem{bjorn}
B.~Garbrecht, 
 Nucl.\ Phys.\ Astropart.\ B 784 (2007) 015.



\bibitem{us}
  A.~Basboll, D.~Maybury, F.~Riva and S.~M.~West,
  Phys.\ Rev.\ D {\bf 76}, 065005 (2007).


\bibitem{me}
  A.~Basboll,
Phys.\ Rev.\ D {\bf 78}, 023528, (2008).

\bibitem{xflat}
J.~McDonald and O.~Seto,
 J.\ Cosmol.\ Astropart.\ Phys. 0807 (2008) 015.



\bibitem{Allahverdi:2008}
  R.~Allahverdi and A.~Mazumdar,
Phys.\ Rev.\ D {\bf 78}, 043511 (2008).
 
\bibitem{GuiMetRioRiva}
 G.~F.~Guidice, L.~Mether, A.~Riotto, F.~Riva,
Phys.\ Lett.\ B\  {\bf 664}, (2008).


\bibitem{Olive08}
 A.~E.~Gumrukcuoglu, K.~A.~Olive, M.~Peloso, M.~Sexton,
Phys.\ Rev.\ D {\bf 78}, 063512 (2008).

\bibitem{RioRiva}
 A.~Riotto,  F.~Riva,
Phys.\ Lett.\ B\  {\bf 670}, (2008).

\bibitem{AllDutMaz}
 R.~Allahverdi, B.~Dutta, A.~Mazumdar, 
Phys.\ Rev.\ D {\bf 78}, 063507 (2008).

\bibitem{brand}
F-Y.~Cyr-Racine, R.~H.~Brandenberger,
J.\ Cosmol.\ Astropart.\ Phys. 0902 (2009) 022.


\bibitem{warsaw}
A.~Kaminska and P.~Pacholek,
arXiv:0901.0478v1.

\bibitem{dufaux}
J-F.~Dufaux,
Phys.\ Rev.\ Lett. {\bf 103}, 041301 (2009).
 
\bibitem{ShoeKuse}
 I.~M.~Shoemaker, A.~Kusenko,
Phys.\ Rev.\ D {\bf 80}, 075021 (2009).



\bibitem{Gumruk}
 A.~E.~Gumrukcuoglu,
arXiv:0910.0854v1


\bibitem{Aitchison}
I.~J.~R.~Aitchison,
arXiv:hep-ph/0505105

\bibitem{mitkatalog}
  A.~Basboll,
  arXiv:0910.0244

\bibitem{ABParis}
  A.~Basboll,
  arXiv:0911.2613


\end{thebibliography}
\end{document}